\newcommand{\lp}{\left(}
\newcommand{\rp}{\right)}

\input{aipcheck}

\documentclass[
    ,final            
  ]
  {aipproc}

\layoutstyle{6x9}

\begin{document}

\title{Parton Distributions with the Combined HERA Charm Production Cross Sections}


\author{Valerio Bertone}{
  address={ Physikalisches Institut, Albert-Ludwigs-Universit\"at Freiburg,\\ 
Hermann-Herder-Stra\ss e 3, D-79104 Freiburg i. B., Germany.}
}

\author{Juan Rojo}{
  address={PH Department, TH Unit, CERN, CH-1211 Geneva 23, Switzerland.\\
$\quad$
\\
Results presented on behalf of the NNPDF Collaboration, \url{http://nnpdf.hepforge.org/}
}
}

\begin{abstract}
Heavy quark structure functions from HERA provide a direct handle on the medium
and small-$x$ gluon PDF. In this contribution,
 we discuss ongoing progress on
the implementation of the FONLL
General-Mass scheme with  running heavy quark masses,
and of its benchmarking with the {\tt HOPPET} and {\tt OpenQCDrad} codes, 
and then
present the impact of the recently released 
combined HERA charm production cross sections in the NNPDF2.3 analysis.
We find that the combined charm data contribute to
constraining the gluon and quarks at small values of Bjorken-$x$.
\end{abstract}

\maketitle


\paragraph{Charm structure function data and PDF fits}

Charm production in deep-inelastic scattering is directly
sensitive to the gluon PDF. The ZEUS and H1 collaborations at the 
HERA collider have measured charm production in DIS with a
wide variety of techniques, and these $F_2^c$ data is included
in all modern PDF fits (see~\cite{Ball:2012wy} for a recent overview).
The HERA experiments have recently released their
combined  data on charm production 
cross sections from HERA Runs I and II~\cite{Abramowicz:1900rp},
 where  a common
consistent data set with the full correlation matrix is provided,
and in addition the combination procedure yields systematic
errors rather smaller than what one would  expect from
the naive combination of all the data, because of the
mutual cross-calibration between H1 and ZEUS.

In this contribution we explore the impact of replacing the separated
H1 and ZEUS $F_2^c$ data with the combined charm 
production cross sections $\widetilde{\sigma}_{\rm NC}^{c}$ in the
NNPDF2.3 analysis~\cite{Ball:2012cx}. We also discuss how one
can generalize the FONLL General-Mass VFN scheme to include
running heavy quark masses in the $\overline{\mbox{MS}}$ scheme.
Heavy quark structure functions in the $\overline{\mbox{MS}}$ scheme lead
to a better behaved perturbative expansion than pole masses~\cite{Alekhin:2010sv} and allow to compare the value of the heavy quark masses used with those
determined by other experiments ({\it e. g.} by LEP data).

\paragraph{FONLL Structure functions in the  $\overline{\mbox{MS}}$ scheme}

Treating heavy quark structure functions in the  $\overline{\mbox{MS}}$ scheme
results in an improved convergence of the perturbative expansion and allows to consistently
 compare
the mass of the charm quark used in the PDF analysis with other determinations.
The NNPDF fits are based on the FONLL GM-VFN scheme for heavy
quark structure functions~\cite{Forte:2010ta}, with 
pole masses as default. It can be shown that  FONLL can be extended
to use  $\overline{\mbox{MS}}$ heavy quark masses,
we discuss here some progress in the this direction. Let us recall that
FONLL can combine different perturbative orders for massive
and massless structure functions: FONLL-A combines the
 NLO 
massless with $\mathcal{O}\lp \alpha_s\rp$ massive pieces, 
FONLL-C achieves the same
at  $\mathcal{O}\lp \alpha_s^2\rp$ and FONLL-B is an intermediate
case where the  $\mathcal{O}\lp \alpha_s^2\rp$ massive calculation
is combined with the NLO massless result.

The first ingredient to consider, as compared
to pole masses, is the NNLO massless PDF evolution,
which is modified by the scheme transformation between pole and
$\overline{\mbox{MS}}$ running masses (NLO evolution is identical
in the two schemes). In particular, the heavy quark thresholds
require different matching conditions with running masses as
compared to pole masses~\cite{Alekhin:2010sv}.
To benchmark the implementation of the $\overline{\mbox{MS}}$  NNLO PDF evolution
in the NNPDF {\tt FastKernel} framework, we report in Table~\ref{HoppetBench} the relative accuracy against the $x$-space evolution code {\tt HOPPET}~\cite{Salam:2008qg}, that also has
the option to perform PDF evolution with 
running heavy quark masses. This benchmark has been performed using the same
parameters as in the Les Houches PDF
 comparison~\cite{Dittmar:2005ed}. We find excellent agreement
over all the kinematic range.

\begin{table}
\begin{tabular}{|c||c|c|c|c|c|}
\hline
 $x$ &      $\epsilon_{\rm rel}\lp u_v\rp $    &   $\epsilon_{\rm rel}\lp  d_v  \rp $ &  $\epsilon_{\rm rel}\lp     L_+ \rp $ &  $\epsilon_{\rm rel}\lp      c^+  \rp $ &    $\epsilon_{\rm rel}\lp   g \rp $  \\
\hline
\hline
$1.0\cdot 10^{-5}$ & $2.30\cdot 10^{-4}$ & $2.63\cdot 10^{-4}$ & $3.28\cdot 10^{-5}$ & $7.10\cdot 10^{-5}$ & $9.39\cdot 10^{-5}$\\
$1.0\cdot 10^{-3}$ & $1.23\cdot 10^{-4}$ & $9.18\cdot 10^{-5}$ & $6.77\cdot 10^{-5}$ & $8.86\cdot 10^{-5}$ & $1.02\cdot 10^{-4}$\\
$1.0\cdot 10^{-2}$ & $2.63\cdot 10^{-4}$ & $3.12\cdot 10^{-4}$ & $9.06\cdot 10^{-5}$ & $1.59\cdot 10^{-4}$ & $1.30\cdot 10^{-4}$\\
$1.0\cdot 10^{-1}$ & $2.69\cdot 10^{-4}$ & $3.99\cdot 10^{-4}$ & $5.29\cdot 10^{-4}$ & $3.36\cdot 10^{-5}$ & $9.15\cdot 10^{-5}$\\
$3.0\cdot 10^{-1}$ & $2.77\cdot 10^{-5}$ & $2.77\cdot 10^{-5}$                       &  $3.79\cdot 10^{-4}$  & $3.79\cdot 10^{-4}$ & $7.21\cdot 10^{-5}$\\
$7.0\cdot 10^{-1}$ & $1.87\cdot 10^{-4}$ & $1.21\cdot 10^{-4}$ & $1.56\cdot 10^{-3}$ & $3.75\cdot 10^{-2}$ & $1.44\cdot 10^{-3}$\\
\hline
\end{tabular}
\caption{\small Relative differences  at $Q^2=10^4$ GeV$^2$ for NNLO PDF evolution in the massless scheme with $\overline{\mbox{MS}}$ running heavy quark masses as implemented in the 
{\tt FastKernel} framework in comparison to  {\tt HOPPET}, $\epsilon_{\rm rel}\equiv |\lp q_i^{\rm fk}(x,Q^2)-q_i^{\rm hop}(x,Q^2)\rp /q_i^{\rm fk}(x,Q^2)|$, for various PDF flavor combinations.
The PDFs and the settings of the comparison are the same as in the
Les Houches benchmark comparisons~\cite{Dittmar:2005ed}.
\label{HoppetBench}}
\end{table}

A second ingredient of the FONLL $\overline{\mbox{MS}}$ masses implementation is the comparison with the  {\tt OpenQCDrad} code\footnote{\url{http://www-zeuthen.desy.de/~alekhin/OPENQCDRAD/}},
which provides charm structure functions in the $N_f=3$ FFN scheme
both for pole masses and for $\overline{\mbox{MS}}$ running masses.
The comparison is done as follows. First of all, we compute
charm structure functions in the massive scheme with pole masses set to
$M_c=\sqrt{2}$ GeV in both codes, and check that there
is reasonable agreement. Then we  transform the pole mass to 
the $\overline{\mbox{MS}}$ running mass,
which is this case corresponds to $m_c(m_c)=1.06$ GeV, and use this as input
of the $\overline{\mbox{MS}}$ running mass computation both
in {\tt FastKernel} in the FFN scheme and in {\tt OpenQCDrad}.
In Table~\ref{f2cbenchopenqcdrad} we present the percentage
difference for charm structure function $F_2^c$ 
in the $\overline{\mbox{MS}}$ scheme, we find good agreement over a wide kinematical range. 

With these two ingredients, it is possible to generalize FONLL
to include running heavy quark masses. Of course the difference between
pole and running masses appears only at  $\mathcal{O}\lp \alpha_s^2\rp$, 
so FONLL-A will
not be changed, while both  FONLL-B and  FONLL-C will be modified
by the scheme transformation.
 A more detailed discussion of the
implementation of running heavy quark masses in FONLL, and
the corresponding impact on parton distributions, will
be presented elsewhere.

\begin{table}[t]
\begin{tabular}{|c||c|c|c|}
\hline
$x \backslash Q^2$ & $\epsilon_{\rm rel}(Q^2=10$ GeV$^2$)  & $\epsilon_{\rm rel}(Q^2=100$ GeV$^2$) &  $\epsilon_{\rm rel}(Q^2=1000$ GeV$^2$)\\
\hline
\hline
$10^{-4}$ &$\qquad$ $1.8\%$  $\qquad$   &$\qquad$ $1.8\%$ $\qquad$  & $\qquad$ $1.1\%$ $\qquad$ \\
$10^{-3}$ &$\qquad$  $0.1\%$ $\qquad$   &$\qquad$ $0.1\%$ $\qquad$ & $\qquad$ $0.4\%$ $\qquad$ \\
$10^{-2}$ &$\qquad$  $0.2\%$ $\qquad$  & $\qquad$ $0.1\%$ $\qquad$ & $\qquad$ $0.8\%$ $\qquad$ \\
\hline
\end{tabular}
\caption{\small Relative differences between the {\tt FastKernel}
implementation  of the charm structure function $F_2^c$
with  $\overline{\mbox{MS}}$ heavy quark running masses
 and the {\tt OpenQCDrad} results, 
$\epsilon_{\rm rel}\equiv |\lp F_{2c}^{\rm fk}(x,Q^2)-F_{2c}^{\rm oqr}(x,Q^2)\rp /F_{2c}^{\rm fk}(x,Q^2)|$. The computation has been
performed in the massive scheme at order $\mathcal{O}\lp \alpha_s^2\rp$ 
with the LH toy PDFs.\label{f2cbenchopenqcdrad}}
\end{table}

\begin{figure}[t]
  \includegraphics[width=.43\textwidth]{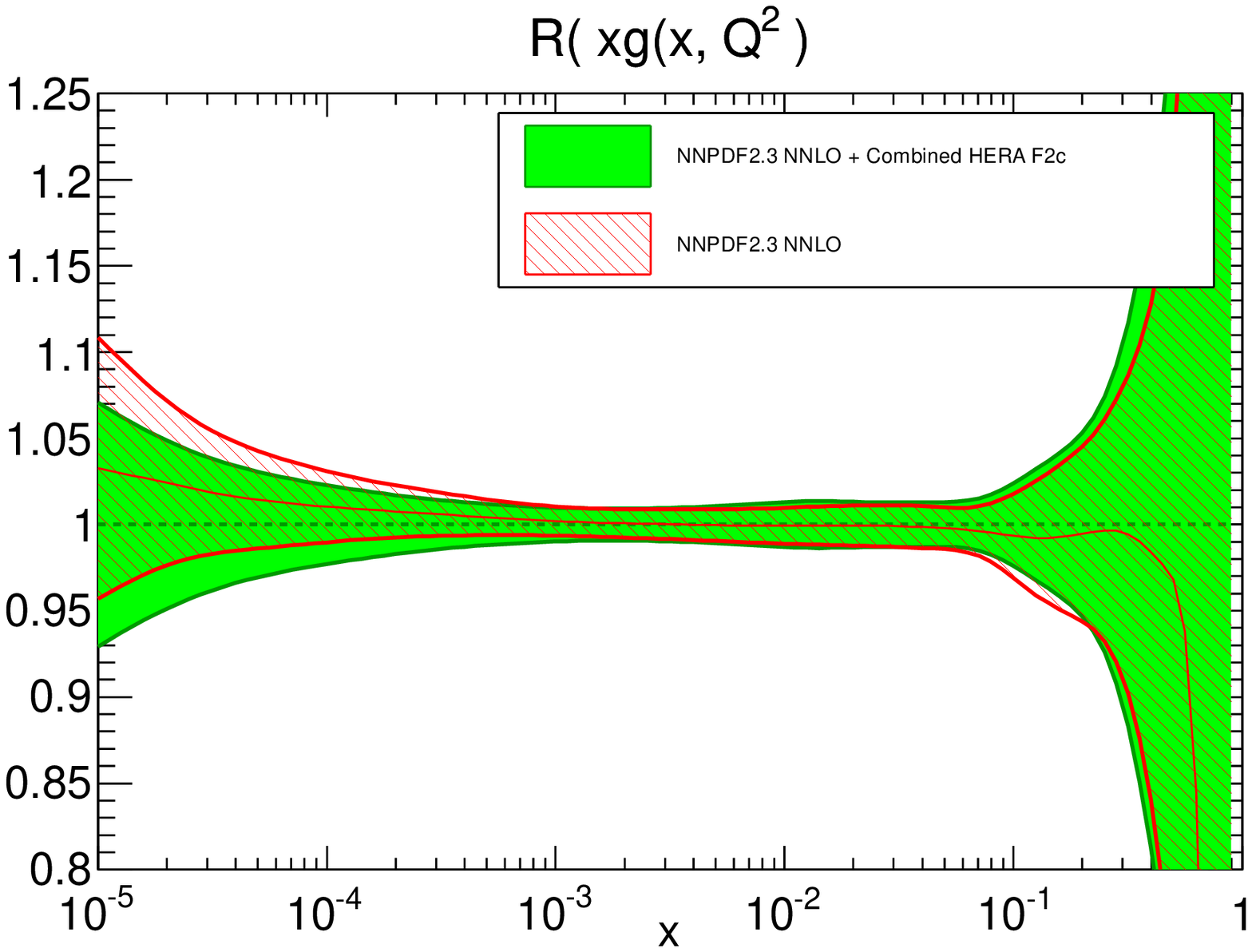}
  \includegraphics[width=.43\textwidth]{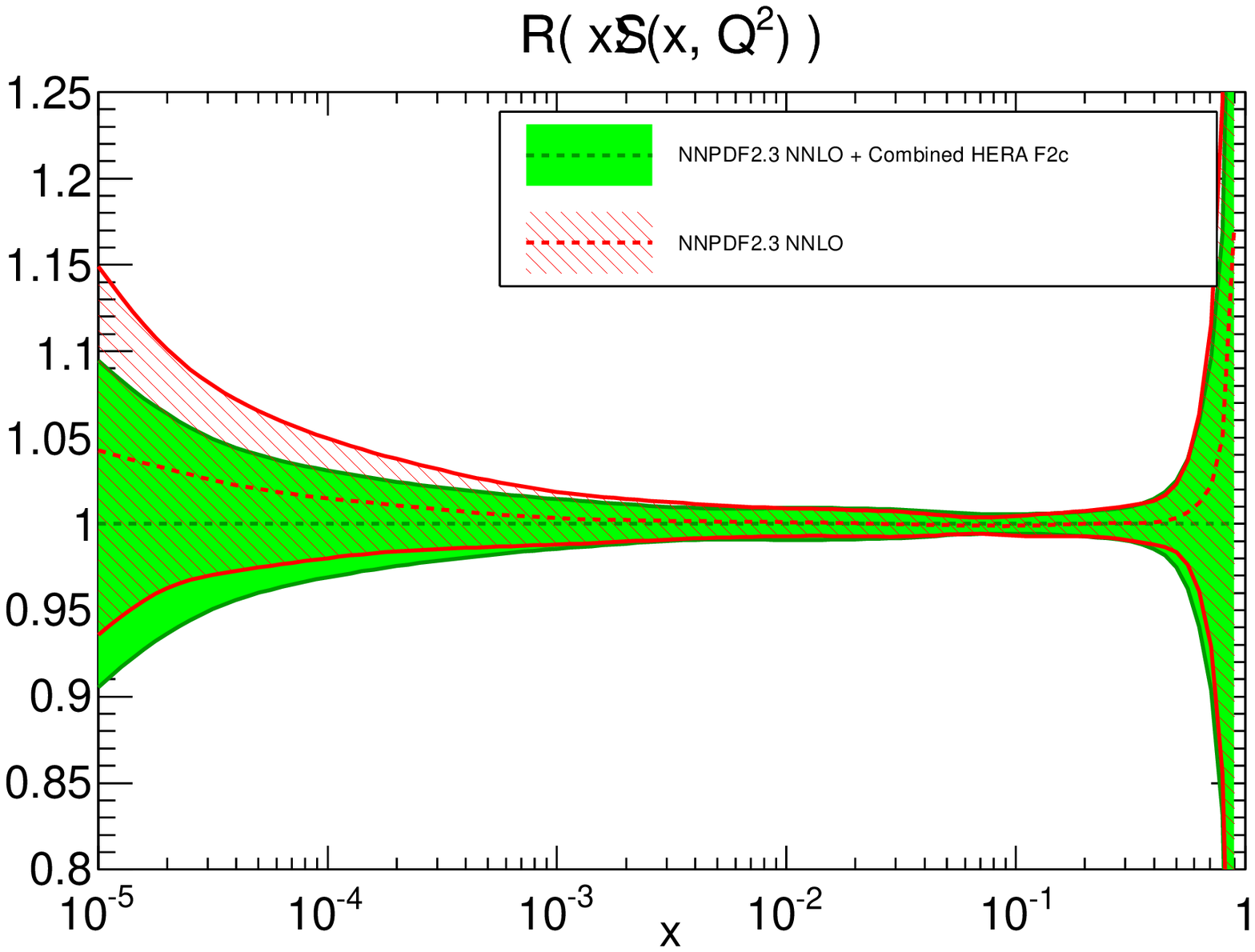}
  \caption{The ratio of the gluon (left plot) and quark singlet (right plot)
between the NNPDF2.3 PDFs and the fit where with the
separated $F_2^c$ H1 and ZEUS data have been
replaced by the combined   $\widetilde{\sigma}_{\rm NC}^{c}$ charm 
reduced cross sections. PDFs have been evaluated at a typical LHC scale of
$Q^2=10^4$ GeV$^2$. \label{fig:pdfs} }
\end{figure}

\paragraph{The impact of the combined HERA charm production data}

The NNPDF analysis have included all the available H1 and ZEUS
charm structure function data since 
NNPDF2.1~\cite{Ball:2011mu,Ball:2011uy,Ball:2011gg,Ball:2012cx}.
We have now implemented the combined HERA 
charm production cross sections
 $\widetilde{\sigma}_{\rm NC}^{c}$  in the NNPDF code,
and performed various fits, at NLO and NNLO, with the same
settings as NNPDF2.3 but replacing the separated H1 and ZEUS
 $F_2^c$ data with the  combined HERA  
 data. We take fully into account
the 43 sources of correlated systematics of the combined
dataset, which include normalization and procedural
uncertainties.
To ease the comparison,
structure functions are computed in the pole mass scheme
with the same mass values as in NNPDF2.3. We obtain
a good description of the data, with a $\chi^2$ per
data point of about $1.2$.

\begin{figure}[t]
  \includegraphics[width=.86\textwidth]{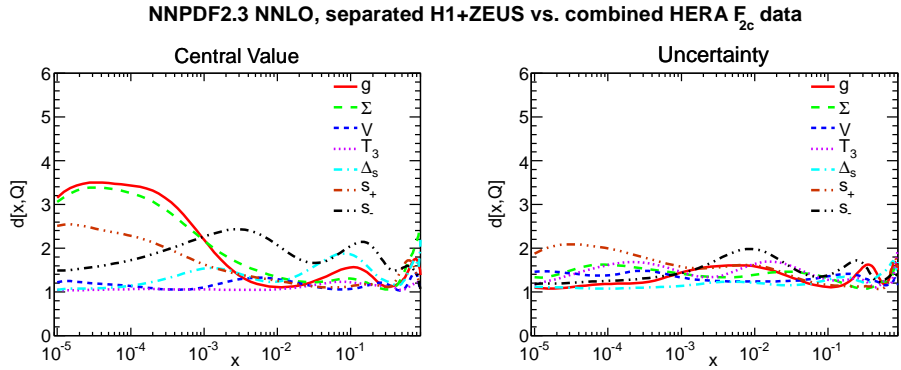}
  \caption{Distances between the two fits of
Fig.~\ref{fig:pdfs}. See~\cite{Ball:2010de} for the definitions.
\label{fig:distances}
}
\end{figure}

We show the impact of the combined HERA charm
data on NNPDF2.3 NNLO in Fig.~\ref{fig:pdfs}  and we also show the 
associated distances
between the two fits in Fig.~\ref{fig:distances}. As we can see,
the net effect of replacing the separated $F_2^c$ data with the combined 
reduced cross sections
is to shift the central value of the small-$x$ gluon and total
quark singlet PDFs by a moderate amount, between one third and
half a sigma of the PDF uncertainty.

\paragraph{Outlook}
We have discussed the implementation of the combined HERA
charm  data in the NNPDF framework,
showed that these new data provide some useful
constrains on the poorly known small-$x$ gluons and quarks, and
reported on the implementation of the FONLL GM scheme with
running heavy quark masses.
We plan to use these results to perform
a determination of $m_c(m_c)$ from the combined HERA charm
production data,
with the same techniques  
used to determine $\alpha_s(M_Z)$~\cite{Lionetti:2011pw,Ball:2011us}.


\begin{theacknowledgments}
We thank M. Corradi and K. Lipka for assistance
with the $F_2^c$ HERA data, and G. Salam for
discussions on {\tt HOPPET}. 
J.~R. is supported by a Marie Curie 
Intra--European Fellowship of the European Community's 7th Framework 
Programme under contract number PIEF-GA-2010-272515.
\end{theacknowledgments}





\end{document}